\documentstyle[prl,aps,multicol,epsfig]{revtex}
\input epsf.tex

\newcommand{\be}{\begin{equation}}
\newcommand{\ee}{\end{equation}}
\newcommand{\ba}{\begin{eqnarray}}
\newcommand{\ea}{\end{eqnarray}}
\newcommand{\ban}{\begin{eqnarray*}}
\newcommand{\ean}{\end{eqnarray*}}

\newcommand{\one}{\leavevmode\hbox{\small1\normalsize\kern-.33em1}}

\begin{document}

\title{Bell-type inequalities for non-local resources}
\author{Nicolas Brunner, Valerio Scarani and Nicolas Gisin}
\address{Group of Applied Physics, University of Geneva, \\ 20 rue de l'Ecole-de-M\'edecine, CH-1211 Geneva 4, Switzerland}
\date{\today}
\maketitle

\begin{abstract}

We present bipartite Bell-type inequalities which allow the two
partners to use some non-local resource. Such inequality can only
be violated if the parties use a resource which is more non-local
than the one permitted by the inequality. We introduce a family of
$N$-inputs non-local machines, which are generalizations of the
well-known PR-box. Then we construct Bell-type inequalities that
cannot be violated by strategies that use one these new machines.
Finally we discuss implications for the simulation of quantum
states.
\end{abstract}

\begin{multicols}{2}

\section{Introduction}

One of the most striking properties of quantum mechanics is
non-locality. It is well known that two separated observers, each
holding half of an entangled quantum state and performing
appropriate measurements, share correlations which are non-local,
in the sense that they violate a Bell inequality \cite{Bell64}.
Indeed this has been demonstrated in many laboratory experiments
\cite{Aspect}. A key feature of entanglement is that it does not
allow the two distant observers to send information to each other
faster than light, i.e. correlation from measurements on quantum
states are no-signaling.

It is an interesting problem to quantify how powerful the
non-local correlations of quantum mechanics are. In order to do
that, one has to use some non-local resource. A quite natural
choice is indeed classical communication. In 2003, Toner and Bacon
showed that a single bit of communication is enough to reproduce
the correlations of the singlet state \cite{Toner}. In the last
years another non-local resource, the PR-box, has also been
proposed to study this problem. Introduced in 1994 by Popescu and
Rohrlich \cite{PR,tsi}, the PR-box was then proven to be a
powerful resource for information theoretic tasks, such as
communication complexity \cite{VanDam,CommCompl} and cryptography
\cite{NLcrypto}. It was also recently suggested that the PR-box is
a unit of non-locality \cite{unitNL}. The PR-box has the appealing
feature that it is intrinsically non signaling, which is of course
not the case of classical communication \cite{1bit}. Note that a
PR-box is a strictly weaker resource than a bit of communication
\cite{Jmod}. Recently, Cerf \textit{et al.} presented a model
using a single PR-box which simulates correlations from any
projective measurement on the singlet \cite{singletPR}. It appears
very natural to extend this study to other quantum states, but
this turns out to be quite difficult, even for non-maximally
entangled pure states of two qubits. In a recent paper we showed a
family of non-maximally entangled states, whose correlations
cannot be reproduced by a single PR-box \cite{Jmod}. In other
words, some non-maximally entangled states require a strictly
larger amount of non-local resources than the maximally entangled
state to be simulated. This suggests that entanglement and
non-locality are different resources. To demonstrate this result
we found a Bell-type inequality allowing some non-local resource;
in this case a single use of a PR-box. Then it was proven that
this inequality is violated by some non-maximally entangled state.

In the present paper, we introduce N-inputs bipartite non-local
machines (NLM), which appear as a natural extension of the
two-inputs PR-box. These machines, denoted $PR_{N}$, have a nice
connection to a family of N-settings Bell inequalities known as
$I_{NN22}$ \cite{dan}, similar to the one that relates the PR-box
to the Clauser-Horne-Shimony-Holt (CHSH) inequality \cite{CHSH}:

\begin{eqnarray} \nonumber
\textrm{CHSH} \quad &\Rightarrow& \quad \textrm{PR-box}  \\
I_{NN22}  \quad &\Rightarrow& \quad PR_{N}    \quad .
\end{eqnarray}

In fact, the structure of the N-inputs NLM can be directly deduced
from the corresponding $I_{NN22}$ inequality. Then we present a
family of N-settings inequalities, $M_{NN22}$, which allow one use
of $PR_{N-1}$ machine. Again, the structure of these new
inequalities is easily deduced from the structure of the
$I_{NN22}$ inequalities, i.e.

\begin{eqnarray}
I_{NN22}  \quad &\Rightarrow& \quad M_{NN22}    \quad .
\end{eqnarray}

Thus a nice construction appears: for any number of settings $N$,
we have a Bell inequality $I_{NN22}$ and the related NLM,
$PR_{N}$, which reaches the upper (no-signaling) bound of the
inequality. Adding one setting we find another inequality,
$M_{(N+1)(N+1)22}$, that cannot be violated by strategies which
require a single use of $PR_{N}$.

The organization of the paper is as follows. In Section II we
present the mathematical tools and introduce the notations by
reviewing the simplest case of two settings on each side. The link
between the PR-box and the CHSH inequality is pointed out. Section
III is devoted to the case of three settings: we introduce a
three-setting NLM and study an inequality for a single use of a
PR-box. In Section IV, the construction of Section III is extended
to the case N settings. Section V concludes the paper by reviewing
the main results about Bell inequalities with and without
resources. Our present work is then clearly situated in this
context.

\section{Tools}

Let's consider a typical Bell test scenario. Two distant
observers, Alice and Bob, share some quantum state. Each of them
chooses between a set of measurements (settings)
$\{A_{i}\}_{i=1..N_{A}}$, $\{B_{j}\}_{j=1..N_{B}}$. The result of
the measurement is noted $r_{A}$, $r_{B}$. Here we will focus on
dichotomic observables and we will restrict Alice and Bob to use
the same number of settings, i.e. $r_{A,B} \in \{0,1\}$ and
$N_{A}=N_{B} \equiv N$. An "experiment" is fully characterized by
the family of $4N^2$ probabilities $P(r_{A},r_{B}|A_{i},B_{j})
\equiv P_{ij}(r_{A},r_{B})$ and can be seen as a point in a
$4N^2$-dimensional probability space . As probabilities must
satisfy

\begin{enumerate}

\item Positivity: $P_{ij}(r_{A},r_{B}) \geq 0 \quad \forall
i,j,r_{A},r_{B}$

\item Normalization: $ \sum_{r_{A},r_{B}=0,1}
P_{ij}(r_{A},r_{B})=1 \quad \forall i,j$

\end{enumerate}

all relevant experiments are contained in a bounded region of this
probability space. Since we want to restrict to no-signaling
probability distributions, we impose also the no-signaling
conditions

\begin{eqnarray}\label{nosignaling}
 \nonumber
\sum_{r_{A}=0,1} P_{ij}(r_{A},r_{B}) &=& P_{j}(r_{B}) \quad \forall i \\
\sum_{r_{B}=0,1} P_{ij}(r_{A},r_{B}) &=& P_{i}(r_{A}) \quad
 \forall j  \quad .
\end{eqnarray}

Conditions (\ref{nosignaling}) mean that Alice output cannot
depend on Bob's setting, and \textit{vice versa}. This shrinks
further the region of possible experiments, and the dimension of
the probability space is now reduced to $d=N(N+2)$. So each
no-signaling experiment is represented by a point in a
$d$-dimensional probability space. In fact the region containing
all relevant probability distributions (strategies), i.e.
satisfying positivity, normalization and no-signaling, form a
polytope, i.e. a convex set with a finite number of vertices. It
is the no-signaling polytope.

One can restrict even further the probability distributions, by
requiring that these are built only by local means, such as shared
randomness. We then obtain a smaller polytope: the local polytope.
The facets of this polytope are Bell inequalities, in the sense
that a probability distribution lying inside (outside) the local
polytope, satisfies (violates) a Bell inequality. The vertices
(extremal points) of this polytope are deterministic strategies
obtained by setting the outputs $r_{A}$ and $r_{B}$ always to 0 or
always to 1. Finding the facets of a polytope knowing its vertices
is a computationally difficult task. In fact, Pitowsky has shown
this problem to be NP-complete \cite{pito}. That's why all Bell
inequalities have been listed for the case of two or three
settings, whereas not much is known for a larger number of
settings.

Let's start with a brief review of the simplest situation: two settings on each side.

This case has been largely studied, and both the local and the
no-signalling polytope have been completely characterized
\cite{Barrett}. The probability space has eight dimensions. We
choose the eight probabilities $P_i(r_A=0)$, $ P_j(r_B=0)$ and
$P_{ij}(r_A=r_B=0)$ to characterize the space.

{\em The local polytope.} The local polytope has 16 vertices. Fine
\cite{fine} showed that all non-trivial facets are equivalent to
the CHSH inequality

\ba \label{chsh} \textrm{CHSH} \equiv
\begin{array}{c|cc} & -1&0\\\hline  -1&1&1\\ 0&1&-1\\
\end{array} \leq 0 \quad.  \ea

Here the notation represents the coefficients that are put in
front of the probabilities, according to \ba
\begin{array}{c|c} & P_i(r_A=0)\\\hline P_j(r_B=0)&
P_{ij}(r_A=r_B=0)\,.
\end{array}\ea

The extremal points (vertices) of the local polytope are
deterministic strategies, i.e. for each setting Alice and Bob
always output 0 or always output 1. Let's do an example: Alice
outputs bit 0 for the first setting $A_{0}$ and outputs 1 for the
second setting $A_{1}$; Bob always outputs 0, for both settings.
This strategy corresponds to the point in probability space

\ba [\underbrace{0_{d},1_{d}}_{\textrm{Alice}};
\underbrace{0_{d},0_{d}}_{\textrm{Bob}}] =
\begin{array}{c|cc} & 1&0\\\hline  1&1&0\\ 1&1&0\\
\end{array} \, .  \ea

All probability distributions lying outside this polytope are
non-local.

The \textit{quantum set} is the set of correlation that can be
obtained by local measurements on quantum states. Inequality
(\ref{chsh}) can indeed be violated by quantum mechanics, and the
maximal violation is $1/\sqrt{2}-1/2\approx 0.2071 $, obtained by
suitable measurements on the singlet state. Of course the quantum
set is included in the no-signaling polytope, but the converse is
not true. There are no-signaling correlations that are more
non-local than those of quantum mechanics. Among these figures
indeed the PR-box.

{\em The no-signaling polytope.}  The no-signaling polytope has 24
vertices: 16 of them are the local vertices seen before and the
eight others are the non-local vertices. Each one of these points
corresponds to a PR-box. Let's make this clear. The PR-box is a
two-inputs, two-outputs NLM. Alice inputs $x$ into the machine and
gets outcome $a$, while Bob inputs $y$ and gets output $b$. The
outcomes are correlated such that $a \oplus b=xy$. The local
marginals are however completely random, i.e.
$P(a=0)=P(b=0)=\frac{1}{2}$, which ensures no-signaling. In
probability space, the PR-box corresponds to the point

\ba\label{PR2} \textrm{PR}  = \frac{1}{2}  \times
\begin{array}{c|cc} & 1&1\\\hline  1&1&1\\ 1&1&0\\
\end{array} \, .  \ea

According to the symmetries $x \rightarrow x+1$, $y \rightarrow
y+1$, $a \rightarrow a+1$, there are eight "equivalent" PR-boxes.
As pointed out in Ref. \cite{Barrett}, there is a strong
correspondence between the eight CHSH facets of the local polytope
and the eight PR boxes. Above each CHSH inequality lies one of the
PR boxes. Each PR-box violates its corresponding inequality up to
$0.5$, which is the maximal value for a no-signaling strategy.
Formally, this correspondence is also pretty obvious by looking at
tables (\ref{chsh}) and (\ref{PR2}). To get the PR-box from the
CHSH inequality, proceed as follows:

\textbf{Recipe.} When the coefficient of a joint probability is
$+1$ or $0$ in the inequality, replace it with $0.5$; when a
coefficient is equal to $-1$, replace it with $0$ in the machine.

In other words when a joint probability appears with a coefficient
$+1$ or $0$, the outputs of the machine are correlated, and when
the coefficient is $-1$, the outputs are anti-correlated. This
simple recipe can be straightforwardly extended to Bell
inequalities with more settings. For a Bell inequality with $N$
settings, we then get a new NLM, denoted $PR_{N}$. This machine
has N inputs and binary outcomes (see Fig. 1).


\begin{center}
\begin{figure}[!h]
\epsfxsize=7cm \epsfbox{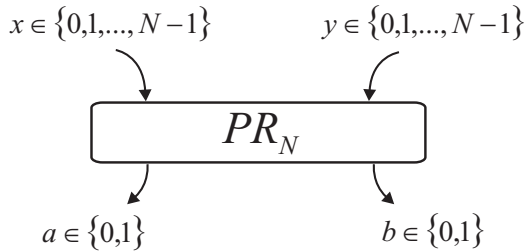} \caption{An N-inputs NLM:
generalization of the two-inputs PR-box.}
\end{figure}
\end{center}

\section{Main result - Three settings}

In this paper we present Bell-type inequalities allowing the use
of some non-local resource. This means that all strategies
satisfying such inequality can be simulated by local means (i.e.
shared randomness, etc) together with some non-local resource ---
for example one NLM. In other words, any strategy violating such
inequality would require a strictly larger amount of non-local
resource than is allowed by the inequality. In the case of two
settings, described in the previous Section, such inequalities
cannot exist. This is because the most elementary non-local
resource, the PR-box, suffices already to generate all the
non-local vertices of the no-signaling polytope.


Therefore we switch to the next case, i.e. three settings (with
two outputs) on each side. Here the situation becomes much more
complicated but remains tractable. All facets of the local
polytope have been listed \cite{dan}. No-signaling strategies are
now living in a 15-dimensional space.

{\em The local polytope.} The local polytope has 64 vertices.
Surprisingly it turns out that each of the $648$ non-trivial
facets is equivalent to one of the two following Bell inequalities

\ba \label{CHSH3}\textrm{CHSH} &\equiv&
\begin{array}{c|ccc} & -1&0&0\\\hline 0&0&0&0\\ -1&1&1&0\\ 0&1&-1&0\\
\end{array} \leq 0  \\
I_{3322} &\equiv& \begin{array}{c|ccc} & -1&0&0\\\hline -2&1&1&1\\ -1&1&1&-1\\ 0&1&-1&0\\
\end{array}\leq 0
\ea

The CHSH inequality is still a facet of the local polytope. This
is a general property of Bell inequalities, known as "lifting"
\cite{lifting}: a facet Bell inequality, defined in a given
configuration, remains a facet when the number of settings,
outcomes or parties is augmented.

Quantum mechanics indeed violates the three-settings CHSH
inequality. The second inequality, $I_{3322}$, is also violated by
quantum mechanics. Furthermore this inequality is relevant, since
it is violated by some quantum states which do not violate the
CHSH inequality \cite{dan}.

{\em The no-signaling polytope.} The local polytope has 72
CHSH-type facets. Above each of these facets lies a PR-box. This
is clear since the CHSH inequality, while still being a facet of
any local polytope with more settings, is a true two-inputs Bell
inequality. Now it is interesting to see that above each
$I_{3322}$ inequality (which is a true three-inputs Bell
inequality) we find a no-signaling strategy which is more
non-local than a PR-box. This strategy is represented by a
three-inputs NLM, defined by the relation $[xy/2]=a+b (\textrm{mod
2})\,$, where $x,y \in \{ 0,1,2 \}$ and $a,b \in \{0,1\}$. This
machine will be refered to as $PR_{3}$. In probability space this
new machine corresponds to the point

\ba\label{PR3} PR_{3} &=&  \frac{1}{2}  \times
\begin{array}{c|ccc} & 1&1&1\\\hline  1&1&1&1\\ 1&1&1&0\\
1&1&0&1\\\end{array} \, .  \ea

Note that $\langle I_{3322}|PR_{3} \rangle =1 $, while $\langle
I_{3322}|PR \rangle =0.5 $ (see Fig. 2). Here we have used a
\textit{scalar product-type} notation $\langle I|S \rangle =z $,
which means that testing inequality $I$ with strategy $S$ gives a
value $z$. The machine $PR_{3}$ can be simply obtained from the
inequality $I_{3322}$ using the Recipe mentioned at the end of the
Section II. One needs two PR-boxes to simulate $PR_{3}$, as shown
in Appendix A. $PR_{3}$ can also be rewritten in the elegant
manner $x=y \, \Leftrightarrow a=b$, which corresponds to the
probability point

\begin{center}
\begin{figure}[b!]
\epsfxsize=7cm \epsfbox{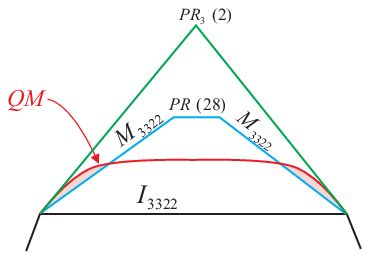} \caption{A facet $I_{3322}$
viewed in a simplified representation of the probability space.
Above the facet, on the hyperplane $I_{3322}=0.5$, lie 28
strategies with a single PR. Above this hyperplane ($I_{3322}=1$)
lie two strategies with a single $PR_{3}$. The local polytope is
in black and the no-signaling in green. In blue is the polytope of
all strategies using at most one PR; $M_{3322}$ is a facet of this
polytope. The quantum set is in red. Note that some quantum states
violate $M_{3322}$ (shaded area). }
\end{figure}
\end{center}

\ba\label{E32}   \frac{1}{2}  \times
\begin{array}{c|ccc} & 1&1&1\\\hline  1&1&0&0\\ 1&0&1&0\\
1&0&0&1\\\end{array} \, .  \ea

The distribution (\ref{E32}) is indeed equivalent to (\ref{PR3})
up to local symmetries: here both Alice and Bob flip their outputs
for their first setting.

In a recent paper Jones and Masanes \cite{lluis} gave a complete
characterization of all the vertices of the no-signaling polytope
for any number of settings and two outcomes --- note that Barrett
\textit{et al.} studied the reversed case: two settings and any
number of outcomes \cite{Barrett}. From their result it is clear
that all vertices of the no-signaling polytope for three settings
and two outputs can be constructed with a $PR_{3}$.

Numerically we find all the vertices of the no-signaling polytope.
We proceed as follows. First we generate all strategies that use
at most one $PR_{3}$. These are all the strategies where Alice and
Bob can choose each of their three inputs in the set
$\{0d,1d,0m,1m,2m,0mf,1mf,2mf\}$. Here $0d,1d$ means that they
deterministically output the value $0$ or $1$; $0m,1m,2m$ means
that they input $0,1,2$ in the machine $PR_{3}$; $0mf,1mf,2mf$
means that they input $0,1,2$ in $PR_{3}$ and flip the output of
the machine. Second, we remove those strategies which are inside
the local polytope by testing all the 648 Bell inequalities.
Finally there are 1344 strategies left which are the non-local
vertices of the (three-inputs two-outcomes) no-signaling polytope.
We find four different classes of those vertices --- given in
Appendix B. A curious feature of those points is that each of them
violates several inequalities of the local polytope. For example
the strategy

\ba\label{PR} PR &=& \frac{1}{2}  \times \begin{array}{c|ccc} & 1&1&0\\\hline  0&0&0&0\\ 1&1&1&0\\
1&1&0&0\\\end{array} \,   \ea

violates the CHSH inequality (\ref{CHSH3}). But it clearly also
violates eight $I_{3322}$-type inequalities, among which

\ba \label{exI3322}
\begin{array}{c|ccc} & -1&0&0\\\hline  -2&1&1&1\\ -1&1&1&-1\\
0&1&-1&0\\\end{array}\leq 0 \quad
\begin{array}{c|ccc} & -1&0&-1\\\hline  0&-1&-1&1\\ -1&1&1&0\\
0&1&-1&0\\\end{array}\leq 0  \,\,.  \ea

Formally this is clear, since each of these eight $I_{3322}$
inequalities (for example (\ref{exI3322})) reduces to the CHSH
inequality (\ref{CHSH3}) once Alice's third setting and Bob's
first setting are discarded. Fig. 3 gives some geometrical
intuition of the situation.

\begin{center}
\begin{figure}[b!]
\epsfxsize=7cm \epsfbox{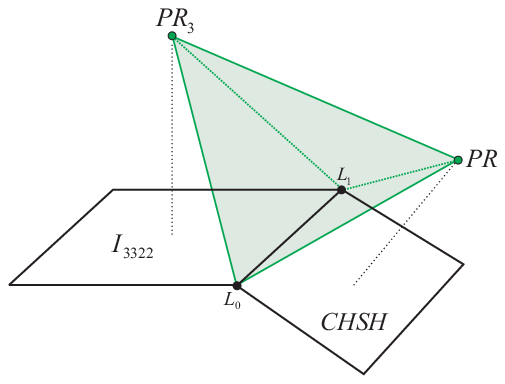} \caption{Simplified
3-dimensional view of a facet of the no-signaling polytope (green
shaded surface). Among the extremal points of this facet are a
PR-box strategy (\ref{PR}), a $PR_{3}$ (\ref{PR3}) strategy, and a
deterministic strategy
$L_{0}=[0_{d},0_{d},0_{d};0_{d},0_{d},0_{d}]$. Behind this facet
is another no-signaling facet which has a vertice
$L_{1}=[1_{d},1_{d},1_{d};1_{d},1_{d},1_{d}]$. Indeed,
$L_{0},L_{1}$ are extremal points of the local polytope. Note that
PR and $PR_{3}$ are both above the CHSH and the $I_{3322}$
facets.}
\end{figure}
\end{center}

{\em Inequality with a PR-box.} We have just seen that, in the
case of three settings on each side, there are two types of NLM,
the PR-box and the $PR_{3}$, generating different types of
non-local vertices of the no-signalling polytope. As  mentioned,
the $PR_{3}$ is a stronger non-local resource than the PR-box ---
it needs two PR-boxes to be simulated. Thus there is a new
polytope, sandwiched between the local and the no-signaling
polytopes. It is formed by all strategies that can be simulated
using at most one PR-box (see Fig. 2). A facet of this polytope
was recently found \cite{Jmod}. It corresponds to the inequality

\ba \label{M3322} M_{3322} &\equiv& \begin{array}{c|ccc}
& -2&0&0 \\\hline  -2&1&1&1\\ -1&1&1&-1\\
0&1&-1&0\\\end{array} \leq 0 \,   \ea

Although $M_{3322}$ is not violated by the maximally entangled
state, it is violated by a family of non-maximally entangled
states of two qubits \cite{Jmod}. Indeed the maximally entangled
state does not violate this inequality, since its correlations can
be simulated using a single PR-box \cite{singletPR}. Note that the
structure of $M_{3322}$ is similar to $I_{3322}$, the only
difference being the coefficient of Alice's first marginal.

We prove now that $M_{3322}$ is a facet of the polytope of all
strategies using at most one PR-box. This result will be extended
to the case of N settings in the next section.


The proof consists of two parts: first we show that no strategy
with a single use of a PR-box violates $M_{3322}$; then we show
that there are (at least) $N(N+2)=15$ linearly independent
strategies using at most one PR-box which saturate $M_{3322}$.
Here we just sketch the idea of the proof, see Appendix C for
details.

To prove the first part, we state a Lemma. Any no-signaling
strategy S violating $M_{3322}$, also violates the two following
inequalities

\ba \nonumber  C_{1} \equiv
\begin{array}{c|ccc} & -1&0&0\\\hline  0&0&0&0 \\ -1&1&1&0\\
0&1&-1&0\\\end{array}\leq 0 \quad C_{2} \equiv
\begin{array}{c|ccc} & -1&0&0\\\hline  -1&1&0&1\\ 0&1&0&-1\\
0&0&0&0\\\end{array}\leq 0  \,\,.  \ea

The proof is straightforward. One needs only to note that, for
no-signaling strategies, joint probabilities are smaller (or
equal) than their respective marginals. Then by inverting the
Lemma, we get the following proposition: if S does not violate
both inequalities $C_{1}$ and $C_{2}$, then S does not violate
$M_{3322}$. Finally it is obvious that with a single PR-box one
can violate either $C_{1}$ or $C_{2}$, but not both at the same
time.

For the second part of the proof, we find numerically eight local
deterministic distributions which saturate $M_{3322}$. Then we
find 57 other strategies with one PR-box saturating $M_{3322}$.
Altogether these strategies form an hyperplane of dimension 14.
This completes the proof that $M_{3322}$ is a facet of the
polytope.

\section{N settings}

In this Section, the results of  Section III are extended to the
case of an arbitrary number of settings $N$. We use a family of
Bell inequalities, known as $I_{NN22}$, which were proven to be
facets of the local polytope \cite{dan}. These inequalities are
generalization of the $I_{3322}$ seen before. For $N$ settings,
the inequality reads

\ba \nonumber  I_{NN22} \equiv
\begin{array}{c|cccccc} & -1 &0&0& \cdots 0&0 \\ \hline
-(N-1)&1&1&1& \dots &1&1
\\ -(N-2)&1& 1&1& \cdots &1&-1 \\
-(N-3)&1& 1&1& \cdots &-1&0 \\
\vdots & \vdots & & & & &\vdots\\
-1 & 1& 1&-1 & \cdots &0& 0 \\
0 &  1&-1 &0& \cdots &0& 0 \\ \end{array}\leq 0 \,.  \ea

Using the Recipe of Section II, we construct a family of
$N$-settings NLM

\ba  PR_{N} \equiv  \frac{1}{2}
\begin{array}{c|cccccc} & 1&1&1& \cdots &1&1 \\ \hline
1&1&1&1& \dots &1&1
\\ 1&1& 1&1& \cdots &1&0 \\
1&1& 1&1& \cdots &0&1 \\
\vdots & \vdots & & & & &\vdots\\
1 & 1& 1&0 & \cdots &1& 1 \\
1 &  1&0 &1& \cdots &1& 1 \\ \end{array}   \,.  \ea

In order to simulate $PR_{N}$ one needs $N-1$ PR-boxes. This is
easily shown using a straightforward generalization of Appendix A.
The inequality

\ba  \label{Mnn22} M_{(N+1)22} \equiv
\begin{array}{c|ccccc} & -N&0& \cdots &0&0 \\ \hline
-N&1&1& \dots &1&1
\\ -(N-1)&1& 1& \cdots &1&-1 \\
-(N-2)&1& 1& \cdots &-1&0 \\
\vdots & \vdots & \vdots & & &\vdots\\
-1 & 1& 1& \cdots &0& 0 \\
0 &  1&-1 & \cdots &0& 0 \\ \end{array}\leq 0   \ea is an
$(N+1)$-setting Bell inequality that cannot be violated by
strategies which require a single use of $PR_{N}$, as proven in
Appendix C. In (\ref{Mnn22}) we have omitted a factor ($N+1$) in
the name of the inequality for practical reasons. Again the
structure of $M_{NN22}$ is similar to $I_{NN22}$, up to Alice's
first marginal: in order to get $M_{NN22}$ from $I_{NN22}$, one
simply changes Alice's first marginal to $-(N-1)$.

So finally we get the following nice construction. For any number
of settings $N$ we have a Bell inequality $I_{NN22}$ and an
$N$-input NLM ($PR_{N}$) which reaches the upper no-signaling
bound of $I_{NN22}$. From there, we construct an $(N+1)$-setting
inequality ($M_{(N+1)(N+1)22}$) which cannot be violated with one
use of $PR_{N}$, i.e.

\begin{equation}
(I_{NN22},PR_{N}) \quad \longrightarrow \quad
(M_{(N+1)(N+1)22},PR_{N+1})
\end{equation}

\section{Conclusion}

To conclude, we review briefly the main results concerning
polytopes and Bell inequalities with and without non-local
resources. We focus on two-outcomes settings. Table I summarizes
the situation. The oldest result is due to Fine, who showed that
all (non-trivial) facets of the two-inputs two-outcomes local
polytope are equivalent to the CHSH inequality \cite{fine}. Then
Collins and Gisin completely characterized the case of three
settings \cite{dan}. In particular they showed that there is a
single new inequality ($I_{3322}$) which is inequivalent to CHSH.
They also found a family of facet inequalities $I_{NN22}$ of the
$N$ setting local polytope, but for $N > 3$ it is not known if
there are other inequalities. The vertices of the no-signaling
polytope for two settings and any number of outcomes have been
characterized by Barrett \textit{et al.} \cite{Barrett}, while
Jones and Masanes studied the reversed case: an arbitrary number
of settings with two outcomes \cite{lluis}.

Not much is known about inequalities allowing non-local resources.
In 2003 Toner and Bacon found inequalities allowing one bit of
communication for the case of two and three settings \cite{bacon}.
They showed that the correlations from measurements on any quantum
state satisfy those inequalities. In the present paper we
introduced a family of $N$ inputs NLMs ($\{ PR_{N} \}_{N \geq
3}$), which are a generalization of the well-known PR-box. These
NLMs can be derived from Bell inequalities in the same way than
the PR-box is derived from the CHSH inequality. Then we presented
a new family of inequalities ($\{M_{NN22} \}_{N \geq 3}$) allowing
one use of $PR_{N}$.

For $N=3$, we get an inequality which cannot be violated with a
single PR-box.  This inequality, presented in a previous work
\cite{Jmod}, is however violated by some non-maximally entangled
state of two qubits. Here we checked numerically that no states of
two qubits violates $M_{4422}$ and $M_{5522}$, which suggests that
these states could be simulated with two PR-boxes, or even a
$PR_{3}$-box. However such model has still not been found.

We acknowledge support from the project QAP (IST-FET FP6-015848).

\section{Appendix A}

In this Appendix, we show how to construct a $PR_{3}$ with 2
PR-boxes.

\begin{center}
\begin{figure}[!h]
\epsfxsize=7cm \epsfbox{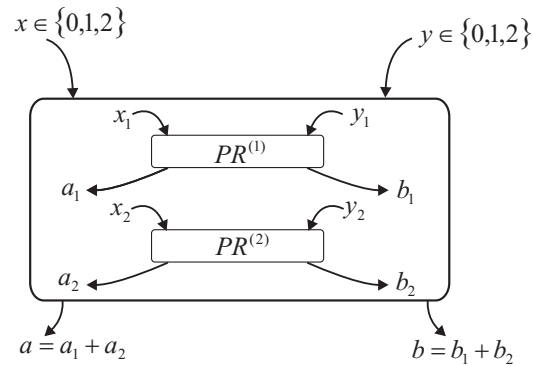} \caption{A $PR_{3}$ with two
PR's. The inputs of each PR, $x_{1,2}$ and $y_{1,2}$ are bits. For
each ternary input $x$ ($y$), there is a combination of $x_{1,2}$
($y_{1,2}$). Finally the output on each side is the sum (modulo 2)
of the two binary outputs of the PR-boxes.}
\end{figure}
\end{center}
Alice and Bob each receive a trit. For each value of the trit they
input one bit in each PR-box. The strategy is the following

\ba
\begin{array}{c|cc}
  x & x_{1} & x_{2} \\ \hline
  0 & 0 & 0 \\
  1 & 0 & 1 \\
  2 & 1 & 0 \\
\end{array}     \quad   \quad
\begin{array}{c|cc}
  y & y_{1} & y_{2} \\ \hline
  0 & 0 & 0 \\
  1 & 1 & 0 \\
  2 & 0 & 1 \\
\end{array}
\ea

where $x,y$ denote the settings, and $x_{i},y_{i}$ are the binary
inputs into PR-box number $i$. Finally, Alice and Bob output the
sum (modulo 2) of both outputs of the PR-boxes. Intuitively the
strategy works as follows. The first machine introduces an
anti-correlation of the outputs for the pair of settings
$x=1,y=2$. The second PR-box does the same for $x=2,y=1$. A nice
way to show that this strategy works is by computing the parity of
the outputs for each pair of settings. So we compute a parity
matrix $P$ by multiplying Alice strategy by the transpose of Bob's
strategy

\ba P= S_{A} S_{B}^{\dag} = \left(
\begin{array}{cc}
   0 & 0 \\
   0 & 1 \\
  1 & 0 \\
\end{array}  \right) \left(
\begin{array}{ccc}
   0 & 1 &0 \\
   0 & 0 &1
\end{array} \right) =  \left(
\begin{array}{ccc}
   0 & 0 &0 \\
   0 & 0 &1  \\ 0&1&0
\end{array} \right)  \quad .
\ea

Note that matrix P has the same structure as the correlation terms
of $I_{3322}$. So Alice and Bob's outputs are identical when a 1
appears in the inequality and different when -1 is in the
inequality.

This construction is easily generalized to $N$ settings. Since
$I_{NN22}$ has $N-1$ correlation terms equal to -1, one simply
uses a PR-box to anti-correlate the outcomes for each of those
terms. Thus it can be shown that a $PR_{N}$ NLM is constructed
with $N-1$ PR-boxes.

\section{Appendix B}

We find four classes of non-local vertices of the three-settings
two-outcomes no-signaling polytope

\ba  S_{1} &=& \begin{array}{c|ccc} & x&x&x\\\hline  x&x&x&x\\ x&x&x&0\\
x&x&0&x\\\end{array} \quad S_{2} = \begin{array}{c|ccc} & x&x&0\\\hline  0&0&0&0\\ x&x&x&0\\
x&x&0&0\\\end{array}  \\
S_{3} &=& \begin{array}{c|ccc} & x&x&0\\\hline  x&x&x&0\\ x&x&x&0\\
x&x&0&0\\\end{array} \quad S_{4} = \begin{array}{c|ccc} & x&x&x\\\hline  x&x&x&x\\ x&x&x&x\\
x&x&0&x\\\end{array} \ea where $x=\frac{1}{2}$. Class $S_{1}$
corresponds to strategies with a $PR_{3}$. They violate maximally
$I_{3322}$, i.e. up to $1$. Classes $S_{2}-S_{4}$ are strategies
which can be obtained with a PR-box. In $S_{2}$, Alice and Bob
have a deterministic output for one of their setting; in $S_{3}$,
only Alice (or Bob) has a deterministic setting; in $S_{4}$, none
outputs deterministic values.

There are 192 vertices in class 1, 288 in class 2, 576 in class 3,
and 288 in class 4. All strategies in the same class violate the
same number of CHSH inequalities and the same number of $I_{3322}$
inequalities. These numbers are summarized in the table below. For
each class of vertices, the number of CHSH and $I_{3322}$
inequalities violated is given.

\begin{center}
\begin{tabular}{||c||c|c||}
 \hline\hline
 Class &    CHSH &    $I_{3322}$  \\ \hline\hline
  $S_{1}$ &  6 & 18  \\
  $S_{2}$  & 1 &  8 \\
  $S_{3}$ &  2& 12 \\
  $S_{4}$ &  4& 24  \\
  \hline\hline
\end{tabular}
\end{center}

\section{Appendix C}

Here it is shown that inequality

\ba \nonumber  M_{NN22} \equiv
\begin{array}{c|cccccc} & -(N-1)&0&0& \cdots 0&&0 \\ \hline
-(N-1)&1&1&1& \dots &1&1
\\ -(N-2)&1& 1&1& \cdots &1&-1 \\
-(N-3)&1& 1&1& \cdots &-1&0 \\
\vdots & \vdots & & & & &\vdots\\
-1 & 1& 1&-1 & \cdots &0& 0 \\
0 &  1&-1 &0& \cdots &0& 0 \\ \end{array}\leq 0 \,,  \ea is a
facet of the polytope containing all strategies that use (at most)
one $PR_{N-1}$. The proof is in two parts. First we show that no
strategy with a $PR_{N-1}$ can violate $M_{NN22}$. Then we show
that the inequality is indeed a facet, i.e. we show that the
strategies saturating the inequality ($M_{NN222}=0$) form a
$(d-1)$-dimensional hyperplane, where $d=N(N+2)$ is the dimension
of the probability space.

\textbf{Part 1.} We start with a Lemma.

{\em Lemma 1.} Let's define the two inequalities

\ba \nonumber  C^{N}_{1} \equiv
\begin{array}{c|cccccc} & -(N-2)&0&0& \cdots 0&&0 \\ \hline
0&0&0&0& \dots &0&0
\\ -(N-2)&1& 1&1& \cdots &1&0 \\
-(N-3)&1& 1&1& \cdots &-1&0 \\
\vdots & \vdots & & & &  &\vdots\\
-1 & 1& 1&-1 & \cdots &0& 0 \\
0 &  1&-1 &0& \cdots &0& 0 \\ \end{array}\leq 0  \ea

\ba \nonumber  C^{N}_{2} \equiv
\begin{array}{c|cccccc} & -(N-2)&0&0& \cdots 0&&0 \\ \hline
-(N-2)&1& 0&1& \dots &1&1\\
-(N-3)&1& 0&1& \cdots &1&-1 \\
-(N-4)&1& 0&1& \cdots &-1&0 \\
\vdots & \vdots & & & & &\vdots\\
0 & 1& 0&-1 & \cdots &0& 0 \\
0 &  0&0 &0& \cdots &0& 0 \\ \end{array}\leq 0 \,.  \ea

Let S be a strategy with N settings for each of the two partners.
S is in a probability space of dimension $N(N+2)$. If S violates
inequality $M_{NN22}$, then S also violates both inequalities
$C^{N}_{1}$ and $C^{N}_{2}$.

{\em Proof.} S violates $M_{NN22}$, i.e.

\ba \label{E1} \nonumber  & & -(N-1)P(A_{0})-
\sum^{N-2}_{k=0}(N-k-1)P(B_{k})+  \sum^{N-1}_{k=0}P(A_{k}B_{0})
 \\ & &
+ \sum_{m=1}^{N-1} \bigg[ \bigg( \sum^{N-m-1}_{k=0}P(A_{k}B_{m})
\bigg) - P(A_{N-m}B_{m}) \bigg] > 0  \ea

According to the no-signaling condition, we have

\ba P(A_{0}) \geq P(A_{0}B_{0})  \ea

\ba (N-1)P(B_{0}) \geq \sum^{N-1}_{k=1}P(A_{k}B_{0}) \ea

\ba P(A_{N-1}B_{1}) \geq 0 \ea

Inserting these relations into (\ref{E1}) we get,

\ban -(N-2)P(A_{0})- \sum^{N-2}_{k=1}(N-k-1)P(B_{k})+
\sum^{N-2}_{k=0}P(A_{k}B_{1}) \\
+\sum_{m=2}^{N-1} \bigg[ \bigg( \sum^{N-m-1}_{k=0}P(A_{k}B_{m})
\bigg) - P(A_{N-m}B_{m}) \bigg]
 > 0 \ean which means S violates inequality $C_{1}^{N}$.

Again from to the no-signaling condition, we have

\ba P(A_{0}) \geq P(A_{0}B_{N-1})  \ea

\ba P(B_{j}) \geq P(A_{1}B_{j}) \quad \quad \forall j \in
\{0,N-1\}\ea

\ba P(A_{1}B_{N-1}) \geq 0 \ea which inserted into (\ref{E1})
gives

\ban -(N-2)P(A_{0})- \sum^{N-3}_{k=0}(N-k-2)P(B_{k})\\ +
\sum^{N-2}_{k=0}P(A_{0}B_{k})  +\sum^{N-1}_{k=2}P(A_{k}B_{0}) \\
+\sum_{m=1}^{N-3} \bigg[ \bigg( \sum^{N-m-1}_{k=2}P(A_{k}B_{m})
\bigg) - P(A_{N-m}B_{m}) \bigg] \\ -P(A_{2}B_{N-2})
 > 0 \ean which means S violates inequality $C_{2}^{N}$.

This completes the first part of the proof.

\textbf{Part 2.} Now we have to show that there are at least
$N(N+2)$ strategies using at most one NLM $P_{N-1}$ on the
hyperplane defined by \ba \label{facet} M_{NN22}=0 \,. \ea

Let's consider only deterministic strategies. We show that there
are $2^{N}$ of them on the $M_{NN22}$ facet.

First we note that there are eight local strategies on the
$M_{3322}$ facet.

\ba  \nonumber
\begin{array}{c|ccc} & 0&1&1\\\hline  1&0&1&1 \\ 0&0&0&0\\
0&0&0&0\\\end{array} \quad
\begin{array}{c|ccc} & 0&1&1\\\hline  0&0&0&0 \\ 0&0&0&0\\
0&0&0&0\\\end{array} \quad
\begin{array}{c|ccc} & 0&1&0\\\hline  0&0&0&0 \\ 1&0&1&0\\
0&0&0&0\\\end{array} \quad
\begin{array}{c|ccc} & 0&1&0\\\hline  0&0&0&0 \\ 0&0&0&0\\
0&0&0&0\\\end{array} \\
\begin{array}{c|ccc} & 0&0&1\\\hline  0&0&0&0 \\ 0&0&0&0\\
1&0&0&1\\\end{array} \quad
\begin{array}{c|ccc} & 0&0&1\\\hline  0&0&0&0 \\ 0&0&0&0\\
0&0&0&0\\\end{array} \quad
\begin{array}{c|ccc} & 0&0&0\\\hline  0&0&0&0 \\ 0&0&0&0\\
1&0&0&0\\\end{array} \quad
\begin{array}{c|ccc} & 0&0&0\\\hline  0&0&0&0 \\ 0&0&0&0\\
0&0&0&0\\\end{array}    \ea

Obviously the marginals fix entirely a deterministic strategy.
Then it is clear that if a three settings strategy

\ba S =
\begin{array}{c|ccc} & A_{0}&A_{1}&A_{2}\\\hline  B_{0}& & & \\ B_{1}& & \cdots & \\
B_{2}& & & \\\end{array}  \ea is on the facet $M_{3322}$, then
both (four settings) strategies

\ba S' =
\begin{array}{c|cccc} & A_{0}&A_{1}&A_{2}& 0\\\hline 0&0&0&0&0 \\ B_{0}& & & &0 \\ B_{1}& & \cdots & &0\\
B_{2}& & & &0 \\\end{array} \\
S'' = \begin{array}{c|cccc} & A_{0}&A_{1}&A_{2}& 1\\\hline   B_{0}& & & & \beta_{0} \\ B_{1}& & \cdots & & \beta_{1} \\
B_{2}& & & & \beta_{2} \\0&0&0&0&0 \\\end{array} \ea are on the
facet $M_{4422}$. The notation $\beta_{j}$ for some correlation
coefficients means that their value depends on Bob's marginal.
Indeed all these strategies are extremal since they are
deterministic.

Then the argument is extended to the next case: for each of the 16
strategies (constructed above) which lie on $M_{4422}$, there are
two strategies on $M_{5522}$. Thus there are $2^N$ deterministic
strategies on $M_{NN22}$. Note that $2^N
> N(N+2)$ for $N \geq 6$. In this case the number of local
strategies on the hyperplane $M_{NN22}=0$ is larger than the
dimension of the probability space. This shows that $M_{NN22}$ is
a facet of the polytope of all strategies using at most one
$PR_{N-1}$. For the case of $N=4$ and $N=5$, we checked
numerically that all strategies using at most one $PR_{N-1}$ form
a subspace of dimension $d-1$, where $d=N(N+2)$ is the dimension
of the probability space.

\end{multicols}

 \newpage

\begin{center}
\begin{table} \label{mainRES}
\begin{tabular}{||c||c|c|c||}
 Resource & $N=2$ & $N=3$ & $N \geq 4$ \\ \hline\hline
  lhv &  CHSH \, \cite{fine} & CHSH+$I_{3322} \, \cite{dan}$ & CHSH+$\{I_{NN22}\}_{N \geq 3}$ +?? \, \cite{dan} \\
  PR-box  & --- & $M_{3322}$ \, \cite{Jmod} & $M_{3322}+$?? \\
  1 bit & --- & \cite{Toner} & ?? \\
  $PR_{N-1}$ & --- & \cite{Jmod} & $M_{NN22}$ \, (this paper) \,
  +??\\ \hline
  no-signaling  & PR \cite{Barrett} & $PR+PR_{3}$ \, \cite{lluis} & $ PR+\{ PR_{N} \}_{N \geq 3}$ \cite{lluis} \\
\end{tabular}
\caption{Main results about Bell inequalities with and without
non-local resources. We consider only the case of binary outcomes.
$N$ is the number of settings. The last line represents the
vertices of the no-signaling polytope. The first column is almost
empty since, for two settings, any no-signaling correlation can be
generated with a PR-box. The question marks mean that it is not
known if there are more of these inequalities. For $N=3$, we are
quite confident that $M_{3322}$ is the only inequality for one use
of a PR-box, though we could not prove it rigorously.}
\end{table}
\end{center}

\end{document}